# Green Hydrogen under Uncertainty: Evaluating Power-to-X Strategies Using Agent-Based Simulation and Multi-Criteria Decision Framework

Frederik Wagner Madsen[1][0009-0003-0663-9817], Joy Dalmacio Billanes[1][0000-0003-2665-9666] Bo Nørregaard Jørgensen[1][0000-0001-5678-6602], and Zheng Ma[1*][0000-0002-9134-1032]

[1] SDU Center for Energy Informatics, The Maersk Mc-Kinney Moller Institute, The Faculty of Engineering, University of Southern Denmark, Odense, Denmark
`{frm, joydbi, bnj, zma}@mmmi.sdu.dk`

**Abstract.** The transition toward net-zero energy systems requires scalable and cost-effective deployment of Power-to-X technologies, particularly green hydrogen production. Despite increasing investments, a critical research gap remains in dynamically assessing how different operational strategies affect the feasibility of hydrogen production under real-world energy market conditions. Most existing studies rely on static, techno-economic models and overlook actor interactions, infrastructure limitations, and regulatory complexity. This paper presents a novel modeling framework that integrates agent-based simulation with multi-criteria decision-making to evaluate green hydrogen production strategies using co-located wind and solar generation. Three operational strategies—grid-only, on-site-only, and hybrid—are applied across three electrolyzer capacity levels (10 MW, 50 MW, and 100 MW) within a Danish case study. Real electricity tariffs, emissions factors, and market data are used to simulate technical, economic, and environmental performance indicators. The results show that hybrid strategies consistently outperform grid-only configurations in terms of cost and emissions while maintaining stable hydrogen output. Although on-site-only strategies minimize emissions and costs, they fail to meet fixed production demands. This framework offers novel scientific contributions by modeling dynamic actor interactions and integrating system performance evaluation into strategic planning. Practically, it provides actionable insights for energy planners and policymakers designing resilient and efficient Power-to-X systems in renewable-rich contexts.

## 1   Introduction

The global shift toward climate neutrality demands deep decarbonization of sectors heavily reliant on fossil fuels. In 2022, the energy sector alone was responsible for nearly half of global greenhouse gas emissions, necessitating urgent transformations in energy generation and consumption patterns[1]. While electrification and energy efficiency improvements are foundational to this transition, certain sectors—such as heavy industry, long-distance transport, and chemicals—are difficult to electrify directly. In this context, Power-to-X (PtX) technologies have emerged as a key enabler of sustainable energy systems, facilitating the conversion of renewable electricity into storable and versatile energy carriers like hydrogen, ammonia, and methanol [2, 3].

PtX offers critical benefits for integrating variable renewable energy (VRE) sources by enabling sector coupling and enhancing energy system flexibility [4, 5]. Among PtX pathways, green hydrogen production through electrolysis has gained significant attention due to its potential to decarbonize industrial processes and fuel markets. Denmark stands at the forefront of this transition with ambitious targets: a 70% reduction in emissions by 2030, 100% renewable electricity, and 4–6 GW of electrolyser capacity supported by strategic government funding [6, 7]. Flagship projects like Kassø and GreenLab Skive illustrate Denmark's leadership in integrating PtX into national energy infrastructure [8, 9].

Despite this momentum, major challenges hinder the effective deployment of PtX systems. These include high capital and operational costs, the need for advanced control strategies under renewable variability, and immature market mechanisms for hydrogen and e-fuels [14]. Additionally, PtX systems are technically complex, requiring seamless integration of distributed energy resources, electrolyzers, storage systems, and grid interaction. The economics of PtX projects are highly sensitive to electricity prices, tariffs, and operational strategies—factors that vary significantly depending on system design and policy frameworks[10].

Current literature primarily focuses on techno-economic assessments and energy system optimization of PtX configurations under static or idealized assumptions[7]. However, there is a lack of dynamic, agent-based modeling frameworks that simulate real-world PtX operation within evolving energy markets and tariff regimes, particularly at the scale of industrial on-site production. Moreover, limited research exists on how different operational strategies—such as grid reliance, hybrid sourcing, or exclusive on-site generation—impact the feasibility of hydrogen production across technical, economic, and environmental dimensions.

To address these gaps, this study presents an integrated simulation and decision-support framework that evaluates the feasibility of on-site green hydrogen production under varying operational strategies. The framework combines:
- Agent-based modeling (ABM) using AnyLogic to simulate the interactions among actors within a PtX ecosystem (e.g., PtX operators, electricity grid, renewable assets, transportation),
- Multi-Criteria Decision-Making (MCDM) methods (PROMETHEE, TOPSIS, VIKOR) to evaluate and rank the performance of different strategies based on key performance indicators (KPIs) such as production cost, emissions, grid usage, and storage needs.

A Danish case study simulates a renewable energy site comprising a 160-hectare PV installation and wind turbines, exploring three electrolyzer capacities (10, 50, and 100 MW) under three operational strategies: Full grid reliance (S1), Full on-site production (S2), and Hybrid approach (S3). Furthermore, the simulation results are assessed using

MCDM tools to determine optimal configurations that balance cost, emissions, and infrastructure demands.

This work contributes to the literature by offering a novel dynamic modeling framework for PtX operational planning and decision-making, grounded in real-world tariff data and market mechanisms. It provides actionable insights for policymakers, planners, and investors navigating the complexity of large-scale hydrogen deployment in renewable-rich contexts.

## 2　Literature Review

### 2.1　Overview of Power-to-X Technologies and Their Role in Energy Transition

Power-to-X (PtX) technologies convert surplus renewable electricity into storable and transportable energy carriers such as hydrogen, methane, methanol, ammonia, and other synthetic fuels [11]. This conversion facilitates sector coupling across electricity, transport, gas, and heating domains, and is crucial for decarbonizing hard-to-electrify sectors such as heavy industry, shipping, and aviation [6].

Hydrogen production via electrolysis is a cornerstone of the PtX concept [4]. Hydrogen can be used directly or further synthesized into various e-fuels and e-chemicals (e.g., e-methanol, e-ammonia) [1]. These products support compatibility with existing energy infrastructures such as gas pipelines and storage tanks [3]. Moreover, emerging applications extend PtX's relevance to sectors like heat, desalination, and advanced materials (e.g., e-steel, e-carbon fibers) [11].

As Denmark transitions to 100% renewable electricity, PtX offers critical flexibility to integrate high shares of intermittent wind and solar power. PtX systems can act as virtual energy storage, helping mitigate curtailment and grid congestion issues [7]. Additionally, Denmark's policy and industrial context—strong offshore wind potential, progressive regulatory frameworks, and R&D investments—make it a particularly fertile ground for evaluating large-scale PtX integration [6].

Despite its potential, PtX faces several systemic challenges: high capital expenditures, fluctuating renewable supply, grid dependency, and technological immaturity (e.g., efficiency losses and degradation of electrolyzers under dynamic loads) [5]. These limitations underscore the need to explore and compare alternative operational strategies, such as full grid reliance, on-site-only generation, or hybrid sourcing, to identify viable business models under different boundary conditions.

### 2.2　Modeling and Decision Support Approaches for PtX Systems

To support strategic PtX planning, a variety of modeling and simulation tools have been developed. These include techno-economic simulation platforms (e.g., PyPSA, energyPRO), integrated energy system models, and process-level optimization tools that evaluate hydrogen and synthetic fuel production [12].

However, most of these tools assume deterministic conditions and lack the flexibility to capture the heterogeneous roles of system actors or evaluate real-time operational strategies across diverse market and tariff settings. Agent-Based Modeling (ABM), in contrast, enables simulation of complex ecosystems involving distributed actors—such as PtX plant operators, grid entities, and transportation services—and their interactions[13]. This is especially relevant for dynamic PtX configurations where infrastructure use, energy procurement, and hydrogen delivery are interdependent.

Complementing ABM, Multi-Criteria Decision Making (MCDM) methods have been extensively used to support infrastructure planning under uncertainty and conflicting objectives [14]. In the context of PtX systems, MCDM enables stakeholders to compare alternative operational strategies not just based on cost, but also environmental performance, storage requirements, infrastructure utilization, and grid impact.

Methods such as TOPSIS, PROMETHEE, and VIKOR are widely adopted due to their capacity to synthesize trade-offs among multiple quantitative and qualitative performance indicators [15]. These methods have been applied across the energy domain—from evaluating micro gas turbine performance to solar farm site selection and hydrogen storage options [16]. In this paper, MCDM tools are deployed to evaluate nine hydrogen production strategies (three capacities × three strategies) simulated through an ABM framework, aligning the decision process with real-world complexity.

## 2.3  Policy and Market Context for PtX in Denmark

Denmark's ambitious energy transition goals—such as achieving a 70% emissions reduction by 2030 and carbon neutrality by 2050—have positioned the country as a PtX policy and innovation leader [7]. Its national PtX strategy, introduced in 2021, promotes the deployment of 4–6 GW of electrolysis capacity by 2030, backed by a 1.25 billion DKK funding scheme [8].

Denmark's PtX ecosystem is supported by high renewable penetration (over 80% electricity from renewables), mature electricity markets, and favorable regulations that encourage behind-the-meter configurations and tariff exemptions for PtX operators co-located with renewable assets[7]. This legal environment enables flexible operational strategies, such as direct consumption of on-site electricity, that are modeled in this study.

Nevertheless, the economic viability of PtX remains precarious. Electrolysis is electricity-intensive, and grid tariffs can constitute a significant operational cost [6]. Participation in ancillary markets—such as FCR (fast activation time, low energy), aFRR (medium activation time, medium energy), and mFRR (slow activation time, high energy)—has shown potential to reduce Levelised Cost of Methanol (LCoM) and improve Net Present Value (NPV) of hydrogen projects [17]. However, commercial competitiveness is still hindered by the need for substantial price premiums for green fuels [9].

Given these market dynamics and infrastructure demands, assessing the trade-offs between full-grid, hybrid, and on-site-only PtX strategies becomes a critical task. This study simulates these strategies in a Danish context using real tariff data and evaluates them with MCDM tools to identify optimal pathways under realistic constraints.

## 3 Methodology

This study adopts a hybrid methodological approach combining agent-based modeling (ABM) and multi-criteria decision-making (MCDM) to evaluate the feasibility of Power-to-X (PtX) strategies under real-world conditions. Building on established business ecosystem modeling frameworks [18] [19], the ABM enables dynamic simulation of PtX operations by capturing how different sourcing strategies—grid-only, on-site-only, and hybrid—affect system performance over time in response to fluctuating tariffs, renewable availability, and actor interactions. Unlike static models, this approach reveals how operational choices impact hydrogen production reliability, cost-effectiveness, and emissions under variable market and infrastructure constraints.

### 3.1 Agent-Based Simulation Framework

The system under investigation is conceptualized as a business ecosystem comprising multiple interacting entities, each fulfilling distinct roles in hydrogen production, electricity procurement, and distribution logistics. These entities are modeled as software agents with dynamic behaviors and interactions.

The simulation is implemented using the AnyLogic™ multi-method modeling platform [20], which supports discrete-event, system dynamics, and agent-based modeling. Each agent represents a stakeholder in the PtX value chain. The key agents and their core responsibilities applied in the study are:

- PtX Company – Owns the electrolyzer facility; responsible for hydrogen production and financial flows.
- Operation Management System (OMS) – Oversees operational logic including dispatch strategy and storage utilization.
- Electrolyzer – Converts electricity (from grid or RE sources) into hydrogen; modeled with a non-linear efficiency curve.
- On-Site Storage – Buffers hydrogen output to align with transportation schedules and end-user delivery.
- Photovoltaic (PV) and Wind Turbine – Provide on-site electricity based on solar irradiation and wind profiles.
- Transportation Company – Manages hydrogen truck logistics to serve demand.
- Hydrogen Consumer – Purchases fixed daily volumes of hydrogen via bilateral agreements.
- Grid System – Provides electricity when RE is insufficient; includes interactions with:
    - Distribution System Operator (DSO) – Applies capacity-based and time-varying tariffs.
    - Transmission System Operator (TSO) – Manages day-ahead market pricing.
    - DataHub – Supplies real-time price and tariff data to OMS.

The model simulates three electrolyzer capacity levels: 10 MW, 50 MW, and 100 MW, each with three operational strategies (grid-based, on-site only, and hybrid). For each experiment, the model tracks key performance indicators (KPIs), including hydrogen production, energy costs, $CO_2$ emissions, and infrastructure utilization (e.g., storage and trucks).

### 3.2 Operational Scenarios and Strategies

The model executes nine simulation experiments—three per capacity level—summarized as follows:
- Strategy 1 (S1) – Hydrogen is produced using electricity solely purchased from the grid; all on-site RE is sold to the market.
- Strategy 2 (S2) – Hydrogen is produced exclusively using on-site renewable energy (PV + wind); surplus is sold to the grid.
- Strategy 3 (S3) – A hybrid approach: RE is prioritized for hydrogen production; grid electricity is used only when RE is insufficient, and excess RE is sold.

All experiments consider a fixed daily hydrogen demand derived from electrolyzer sizing, assumed at 5,000 full load hours annually. Real 2024 market data on spot prices, $CO_2$ intensities, DSO/TSO tariffs, and efficiency profiles are integrated [21] [22] [23] [24] [25].

The simulation outputs a set of KPIs per experiment:
- Produced hydrogen (tonnes)
- Grid electricity consumption (MWh) and cost (DKK)
- Hydrogen production cost (DKK/kg)
- Electricity sold to grid and revenue
- $CO_2$ emissions (tonnes and kg $CO_2$/kg $H_2$)
- Electrolyzer utilization (hours)
- Storage and truck utilization rates

### 3.3 Multi-Criteria Decision-Making (MCDM) Evaluation

To evaluate the relative performance of each strategy across competing dimensions, the simulation results are post-processed using three well-established MCDM methods [26] [27]:
- TOPSIS – Ranks alternatives based on closeness to an ideal solution.
- PROMETHEE II – Uses pairwise preference comparisons with outranking flows.
- VIKOR – Identifies compromise solutions based on regret and utility measures.

The decision matrix includes 13 KPIs (as criteria), standardized using min-max normalization. Each criterion is classified as either a benefit (e.g., hydrogen production) or a cost (e.g., grid electricity consumption, emissions).

To ensure fairness and robustness, criteria weights are derived using a hybrid method that averages:
- Equal weights – Emphasizing transparency.

- Entropy-based weights – Emphasizing information richness and data variability [15].

This integrated weighting approach balances subjective simplicity with data-driven objectivity.

### 3.4 Evaluation Workflow

The simulation and evaluation workflow consists of the following steps:
1. Model configuration with capacity and strategy inputs.
2. Agent-based simulation over a representative operational period.
3. Extraction and aggregation of performance metrics.
4. Construction and normalization of MCDM decision matrix.
5. Ranking of strategies using VIKOR, PROMETHEE, and TOPSIS.
6. Comparison and interpretation of method outputs to determine robust strategies.

This combined simulation-decision framework provides a scalable, actor-based method for PtX operational planning that is adaptable to evolving market conditions, regulatory changes, and infrastructure constraints.

## 4 Case Study and Scenario Design

### 4.1 Case Overview: Renewable Site and Hydrogen Production Concept

This study investigates the techno-economic feasibility and operational strategies of Power-to-X (PtX) systems for green hydrogen production in Denmark. The selected case simulates a PtX facility co-located with a high-capacity renewable energy site in a Danish context, reflecting the country's strategic investment in decentralized PtX production using wind and solar power.

The case comprises a 160-hectare photovoltaic (PV) installation and four 4.2 MW wind turbines, resulting in a total peak generation capacity of 272.8 MW. This hybrid RE configuration reflects realistic infrastructure scaling for future PtX hubs in Denmark, such as GreenLab Skive and the Kassø PtX cluster, and ensures a representative resource profile for evaluating hydrogen production scenarios.

The facility includes an electrolyzer system powered by both on-site renewables and the grid, with options for intermediate hydrogen storage and delivery by road transport. The operational model considers regulatory, market, and infrastructure constraints, including Danish grid tariffs, real-time electricity prices, and $CO_2$ emissions factors.

### 4.2 Electrolyzer Capacities and Design Assumptions

To assess scalability, the simulation models three electrolyzer capacity levels:

- 10 MW – representing pilot-scale PtX projects;
- 50 MW – medium-scale commercial plants;

- 100 MW – large-scale infrastructure aimed at export markets.

Each capacity scenario is evaluated using a fixed daily hydrogen demand based on 5,000 full-load hours of operation annually, in line with Danish hydrogen production forecasts. The key simulation parameters used across all scenarios, including tariff inputs, hydrogen density, and efficiency assumptions, are summarized in Table 1. Water use and waste heat recovery were excluded from this analysis to focus on electricity-driven process dynamics.

Table 1. Core parameters for electrolyzer case scenarios

| Parameter | Values | Source |
| --- | --- | --- |
| Electrolyzer capacity | 10, 50, 100 MW | Assumed scale |
| Hydrogen demand (daily) | 2,390; 11,952; 23,903 kg/day | Calculated |
| Electricity spot prices | 2024 dataset | [21] |
| $CO_2$ emission factors | 2024 dataset | [21] |
| Grid tariffs (TSO and DSO) | Dynamic for 2024 | [22] [23] |
| Electrolyzer efficiency | Dynamic efficiency curves | [24] |
| Hydrogen LHV | 120 MJ/kg (33.33 kWh/kg) | [25] |

### 4.3 Operational Strategies

Each electrolyzer capacity scenario is simulated using **three operational strategies**, designed to reflect varying levels of energy independence and market interaction.

- Strategy 1 (S1: Grid-Based). The PtX facility sells all on-site renewable energy (RE) to the grid and produces hydrogen using electricity purchased from the grid. This provides a baseline for evaluating grid dependency and associated costs/emissions.
- Strategy 2 (S2: On-Site Only). Hydrogen is produced exclusively using electricity from on-site PV and wind installations. Any surplus RE is sold to the grid. This strategy maximizes local energy utilization and avoids grid procurement costs.
- Strategy 3 (S3: Hybrid). On-site RE is prioritized for hydrogen production. When insufficient, grid electricity is procured to meet demand. Surplus RE is sold to the grid. This hybrid approach balances independence and reliability.

### 4.4 Simulation Parameters and Assumptions

The simulations are implemented in AnyLogic™ using an agent-based framework that models the interactions of key actors, including the PtX company, electrolyzer, storage tanks, grid operators, RE generation units, hydrogen consumers, and transport services.

All scenarios include real-world electricity market data (spot prices, $CO_2$ factors), dynamic efficiency curves for electrolyzers, and time-varying DSO/TSO tariffs for both consumption and production. Each experiment simulates one operational year, with hourly resolution.

The simulation tracks a consistent set of Key Performance Indicators (KPIs), including:

- Hydrogen production (tonnes);
- Grid electricity consumption and cost;
- Hydrogen production cost (DKK/kg);
- Electricity sales revenue;
- $CO_2$ emissions (total and per kg $H_2$);
- Storage and truck utilization rates;
- Electrolyzer full-load operating hours.

### 4.5 Summary of Simulation Experiments

A total of nine simulation experiments were conducted, representing all combinations of three electrolyzer capacities and three operational strategies. These are summarized in Table 2, which outlines the configuration of each experiment in terms of strategy, energy source, and daily hydrogen demand.

Table 2. Scenario–Strategy Matrix for Simulation Experiments

| Experiment ID | Electrolyzer Capacity | Strategy | Electricity Source | Daily Hydrogen Demand |
|---|---|---|---|---|
| 1.1 | 10 MW | S1 | Grid only | 2,390 kg |
| 1.2 | 10 MW | S2 | On-site RE only | 2,390 kg |
| 1.3 | 10 MW | S3 | Hybrid (RE + Grid) | 2,390 kg |
| 2.1 | 50 MW | S1 | Grid only | 11,952 kg |
| 2.2 | 50 MW | S2 | On-site RE only | 11,952 kg |
| 2.3 | 50 MW | S3 | Hybrid (RE + Grid) | 11,952 kg |
| 3.1 | 100 MW | S1 | Grid only | 23,903 kg |
| 3.2 | 100 MW | S2 | On-site RE only | 23,903 kg |
| 3.3 | 100 MW | S3 | Hybrid (RE + Grid) | 23,903 kg |

## 5 Results

This section presents the outcomes of nine simulation experiments assessing the operational performance of three electrolyzer capacities (10 MW, 50 MW, and 100 MW) across three hydrogen production strategies (grid-only, on-site RE-only, and hybrid). Key performance indicators include hydrogen production, electricity consumption and costs, $CO_2$ emissions, and infrastructure utilization. Tables 3–5 present detailed results for each scenario, followed by an integrated MCDM-based ranking.

**Scenario 1: 10 MW Electrolyzer.** Table 3 summarizes the key performance indicators (KPIs) for the 10 MW electrolyzer configuration. Hydrogen production remains equal under Strategy 1 (grid-based) and Strategy 3 (hybrid), while Strategy 2 (on-site RE only) yields a 42% reduction in production due to intermittent resource availability. Grid electricity usage and emissions are highest under Strategy 1, while Strategy 2 achieves zero emissions and the lowest cost per kg of hydrogen. Strategy 3 offers a balance of full hydrogen delivery and reduced emissions and costs.

**Table 3.** Results for Scenario 1 – 10 MW Electrolyzer

| Metrics | 1.1 (S1: Grid-only) | 1.2 (S2: On-site RE only) | 1.3 (S3: Hybrid) |
|---|---|---|---|
| Produced hydrogen [ton] | 868 | 502 | 868 |
| Total grid electricity consumption [MWh] | 49865 | 0 | 20203 |
| Cost of grid electricity consumption [mDKK] | 76 | 0 | 37 |
| Hydrogen production cost [DKK/kg] | 87 | 5 | 42 |
| Total electricity sold to grid [MWh] | 278536 | 248891 | 248891 |
| Revenue from selling electricity [mDKK] | 103 | 92 | 92 |
| CO2 emissions [ton] | 5170 | 0 | 2059 |
| CO2 emissions [kgCO2/kgH2] | 6 | 0 | 2 |
| Electrolyzer usage [full load hours] | 4987 | 2966 | 4987 |
| On-site storage size [kgH2] | 100 | 100 | 100 |
| On-site storage utilization [%] | 16 | 10 | 16 |
| Number of trucks | 2 | 2 | 2 |
| Average hydrogen transportation truck utilization [%] | 56 | 52 | 56 |

**Scenario 2: 50 MW Electrolyzer.** Table 4 presents the KPI results for the 50 MW electrolyzer case. Similar trends are observed: Strategy 1 and Strategy 3 meet full hydrogen demand, while Strategy 2 results in underproduction. Strategy 3 again achieves notable reductions in $CO_2$ emissions and electricity costs compared to full grid dependency. Infrastructure demands increase proportionally with electrolyzer capacity.

**Table 4.** Results for Scenario 2 – 50 MW Electrolyzer

| Metrics | 2.1 (S1: Grid-only) | 2.2 (S2: On-site RE only) | 2.3 (S3: Hybrid) |
|---|---|---|---|
| Produced hydrogen [ton] | 4338 | 1735 | 4338 |
| Total grid electricity consumption [MWh] | 249317 | 0 | 144050 |
| Cost of grid electricity consumption [mDKK] | 378 | 0 | 237 |
| Hydrogen production cost [DKK/kg] | 87 | 5 | 55 |
| Total electricity sold to grid [MWh] | 278536 | 173287 | 173287 |
| Revenue from selling electricity [mDKK] | 103 | 61 | 61 |
| $CO_2$ emissions [ton] | 25848 | 0 | 14668 |
| $CO_2$ emissions [$kgCO_2/kgH_2$] | 6 | 0 | 3 |
| Electrolyzer usage [full load hours] | 4986 | 2105 | 4986 |
| On-site storage size [$kgH_2$] | 1000 | 1000 | 1000 |
| On-site storage utilization [%] | 36 | 21 | 36 |
| Number of trucks | 6 | 6 | 6 |
| Hydrogen transportation truck utilization [%] | 76 | 61 | 76 |

**Scenario 3: 100 MW Electrolyzer.** Table 5 provides the results for the 100 MW electrolyzer configuration. At this scale, Strategy 1 becomes prohibitively expensive and carbon-intensive. Strategy 2 maintains low cost and zero emissions but produces only 31% of the required hydrogen. Strategy 3 successfully combines production reliability with moderate emissions and cost savings. Again, the higher capacity electrolyzer increases the infrastructure need of both on-site storage and number of trucks, however the utilization rate of the infrastructure is better.

**Table 5.** Results for Scenario 3 – 100 MW Electrolyzer

| Metrics | 3.1 (S1: Grid-only) | 3.2 (S2: On-site RE only) | 3.3 (S3: Hybrid) |
|---|---|---|---|
| Produced hydrogen [ton] | 8676 | 2732 | 8676 |
| Total grid electricity consumption [MWh] | 498633 | 0 | 329577 |
| Cost of grid electricity consumption [mDKK] | 757 | 0 | 531 |
| Hydrogen production cost [DKK/kg] | 87 | 5 | 61 |
| Total electricity sold to grid [MWh] | 278536 | 109497 | 109496 |
| Revenue from selling electricity [mDKK] | 103 | 36 | 36 |
| $CO_2$ emissions [ton] | 51695 | 0 | 33637 |
| $CO_2$ emissions [$kgCO_2/kgH_2$] | 6 | 0 | 4 |
| Electrolyzer usage [full load hours] | 4986 | 1690 | 4986 |
| On-site storage size [$kgH_2$] | 2000 | 2000 | 2000 |
| On-site storage utilization [%] | 41 | 25 | 41 |
| Number of trucks | 11 | 11 | 11 |

| Hydrogen transportation truck utilization [%] | 79 | 61 | 79 |
|---|---|---|---|

**Multi-Criteria Evaluation of Strategy-Scenario Combinations.** Using VIKOR, PROMETHEE II, and TOPSIS methods, all nine configurations were evaluated based on 13 normalized KPIs. Figure 1 presents the ranking of the experiments using the three methods, while Table 6 summarizes the average scores. Figure 1 reveals consistency among the three MCDM methods regarding experiment rankings. Both PROMETHEE and the WPM rank Experiment 1.2 as the highest and Experiment 2.2 as the second highest. Conversely, TOPSIS ranks Experiment 2.2 first and Experiment 1.2 second. Examination of the KPIs in Tables 3 and 4 suggests that TOPSIS prioritizes hydrogen production and the utilization of storage and trucks, whereas PROMETHEE and WPM place greater emphasis on cost-related KPIs. Nevertheless, all three methods identify Experiment 3.1 as the least favorable.

From Table 7 it is seen that Strategy 2 consistently outperforms the two other strategies across all electrolyzer size scenarios, whereas Strategy 1 exhibits the lowest performance. Overall, Experiment 1.2 achieves the highest average ranking score of 8.67, closely followed by Experiment 2.2 with a score of 8.33.

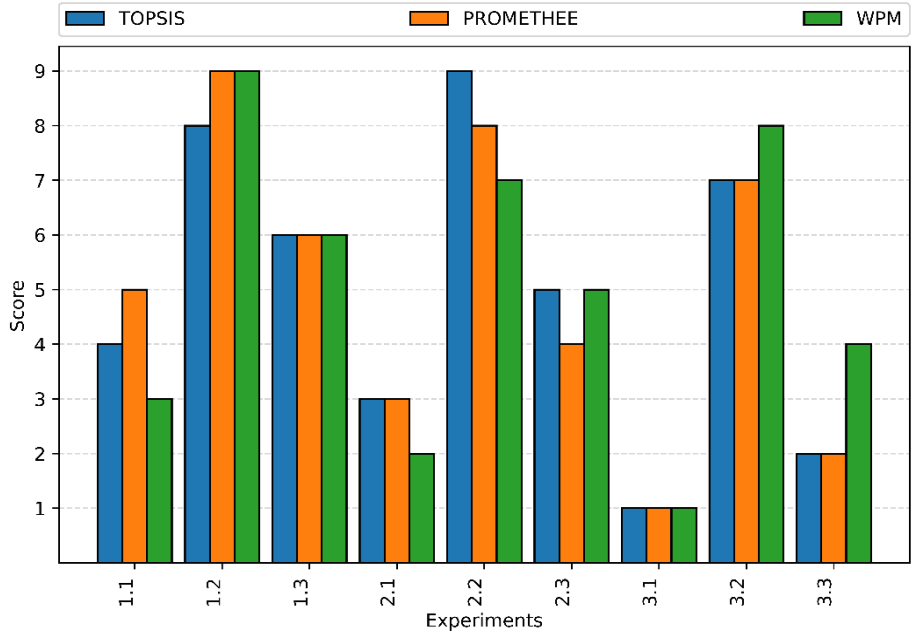

**Fig. 1.** Rankings of strategies under VIKOR, TOPSIS, and PROMETHEE methods

Table 6. MCDM-based ranking of experiments (average of 3 methods)

| Experiment ID | Strategy | Electrolyzer Capacity | Average Score |
|---|---|---|---|
| 1.2 | S2 | 10 MW  | 8.67 |
| 2.2 | S2 | 50 MW  | 8.00 |
| 3.2 | S2 | 100 MW | 7.33 |
| 1.3 | S3 | 10 MW  | 6.00 |
| 2.3 | S3 | 50 MW  | 4.67 |
| 1.1 | S1 | 10 MW  | 4.00 |
| 2.1 | S1 | 50 MW  | 2.67 |
| 3.3 | S3 | 100 MW | 2.67 |
| 3.1 | S1 | 100 MW | 1.00 |

## 6  Discussion

This section interprets the results in the context of existing literature, drawing conclusions regarding the scalability, emissions performance, and economic viability of Power-to-X (PtX) systems in Denmark and similar energy systems.

**Scalability and Production Security.** The consistent underperformance of Strategy 2 (on-site RE only) across all capacity levels demonstrates the practical limitations of relying solely on local renewables for firm hydrogen output. This aligns with [3], that PtX operations must accommodate variability in renewable generation through flexible system design and supplemental power sources. Strategy 3 emerges as the most robust approach, maintaining high hydrogen production levels while reducing dependency on grid electricity. This supports findings by [28] and [14] that emphasized the importance of hybridization and grid-assisted designs in achieving stable PtX operations.

**Emissions and Environmental Implications.** Consistent with [4] and [1], the study demonstrates that grid-based strategies incur significant emissions due to the carbon intensity of electricity, even in relatively clean systems like Denmark's. Strategy 2 eliminates emissions but sacrifices production quantity. Strategy 3 significantly reduces emissions without compromising output, which is critical in scenarios requiring consistent fuel supply to decarbonize hard-to-electrify sectors [29] [5]. This suggests that hybrid PtX systems are essential for meeting environmental goals while preserving system resilience, especially in transition periods before full renewable grid penetration is achieved.

**Economic Viability and Tariff Sensitivity.** Hydrogen production costs under Strategy 1 far exceed those reported in techno-economic benchmarks (e.g., [10]), primarily due to grid electricity prices and DSO/TSO tariffs. In contrast, Strategy 3 offers a feasible cost-performance compromise, aligning with analysis by [6], who reported significant cost advantages in hybrid configurations with dynamic market participation. The advantage of Strategy 3 also aligns with findings by Søgaard-Deakin and Xydis [8], who noted that optimal site location and tariff configurations can significantly enhance PtX economics. By using behind-the-meter RE and selectively accessing grid supply, Strategy 3 leverages regulatory provisions while securing supply.

**Infrastructure Optimization.** The simulation results confirm that larger electrolyzer systems dramatically increase storage and transportation requirements. This finding supports [3] and [11], which stressed the importance of logistics integration in PtX scale-up. The higher utilization rates of infrastructure under Strategy 2 indicate its efficiency in asset use, but this comes at the cost of reduced hydrogen output. Strategy 3 again balances infrastructure use and energy yield.

**Strategic and Policy Implications.** The superiority of hybrid approaches has clear implications for PtX policy design. As shown in this study and in the Danish context outlined by [7] and [22], policies that incentivize on-site generation with partial grid support — such as tariff exemptions or dynamic pricing — can enhance both the cost-efficiency and environmental performance of PtX facilities. From a system planning perspective, this study supports the EU Green Deal's call for regional hydrogen networks by illustrating how localized hybrid plants can anchor flexible PtX clusters, as discussed by [9].

## 7   Conclusion

This study proposed and evaluated an integrated simulation and decision-support framework for assessing Power-to-X (PtX) system operations under varying production strategies and scale levels. By applying agent-based modeling (ABM) and Multi-Criteria Decision-Making (MCDM) methods, the paper explored how different operational configurations—full-grid reliance, exclusive on-site renewables, and hybrid sourcing—impact the techno-economic feasibility, environmental performance, and infrastructure requirements of green hydrogen production in Denmark.

The results demonstrate that while on-site renewable-only strategies offer the lowest emissions and production costs, they consistently fail to meet fixed hydrogen demand across all electrolyzer capacity levels. Grid-based strategies, by contrast, meet production targets but suffer from significantly higher emissions and electricity costs due to their dependence on market-based electricity prices and tariffs. Hybrid strategies emerge as a robust middle ground, achieving full production volumes while considerably reducing emissions and costs compared to grid-only approaches. This strategic balance is further supported by the MCDM results, where hybrid strategies consistently ranked high across multiple performance indicators and decision criteria.

Scientifically, this study contributes a novel combination of ABM and MCDM techniques for dynamic, actor-based PtX system analysis—addressing a critical gap in existing literature, which often relies on static or deterministic techno-economic models. The proposed methodology not only allows for a granular simulation of stakeholder interactions and market constraints but also enhances decision quality by incorporating environmental, operational, and infrastructural trade-offs into an integrated evaluation framework.

Practically, the findings offer concrete insights for PtX developers, policymakers, and system planners. The hybrid model presented here aligns with Denmark's PtX strategy and demonstrates how partial reliance on the grid, combined with behind-the-meter renewable production, can improve cost efficiency while mitigating carbon emissions. This supports ongoing policy efforts related to tariff exemptions, ancillary service market participation, and infrastructure planning for decentralized PtX deployment.

Nonetheless, the study presents several limitations. First, the simulation assumes constant electrolyzer efficiency and does not account for degradation effects, which could influence long-term viability. Second, the scope of KPIs, while comprehensive, excludes considerations such as heat integration and water use, which are critical for full system life-cycle assessment. Third, while MCDM methods capture decision complexity, the weighting approach—though hybridized—may still introduce subjective bias depending on the use case.

To address these limitations, future work should extend the model to include component degradation, water consumption, and thermal energy recovery. Integration with carbon capture and e-fuel synthesis pathways would also enhance relevance for sector coupling scenarios. Furthermore, additional MCDM sensitivity analysis or stakeholder-informed weighting schemes could improve robustness and applicability in participatory planning contexts. A regional-scale extension of the simulation framework—such as linking multiple PtX hubs or evaluating national PtX deployment strategies—would offer further value for system-level optimization and investment planning.

In summary, this paper advances both methodological and applied understanding of PtX operations in renewable-rich energy systems. The proposed framework provides a scalable foundation for evaluating real-world PtX strategies under dynamic and uncertain market conditions, and supports the responsible, data-driven deployment of green hydrogen infrastructure across Denmark and beyond.

**Acknowledgments.** This paper is part of the project "INNOMISSION II: MissionGreenFuel-Digitalization and test for dynamic and flexible operation of PtX components and systems (DYNFLEX)" funded by Innovation Fund Denmark; Part of the project titled "Danish Participation in IEA IETS Task XXI - Decarbonizing industrial systems in a circular economy framework", funded by EUDP (project number: 134233-511205); Part of the project titled "Danish participation in IEA IETS Task XXII - Climate Resilience and Energy Adaptation in Industry under Uncertainty", funded by EUDP (project number: 95-41006-2410288).

Disclosure of Interests. The authors have no competing interests to declare that are relevant to the content of this article